\documentclass[12pt]{article}
\usepackage{cite,epsfig,amssymb,amsmath,graphicx,color,arydshln}
\begin{document}

\baselineskip=.22in
\renewcommand{\baselinestretch}{1.2}
\renewcommand{\theequation}{\thesection.\arabic{equation}}
\renewcommand{\thefootnote}{\fnsymbol{footnote}}

\vspace{10mm}

\begin{center}
{{\Large \bf Constraints on Flows in Ho${\check {\bf r}}$ava-Lifshitz
Gravity\\[3mm]
by Classical Solutions}\\[17mm]

Taekyung Kim \footnote{Electronic address: {\tt pojawd@skku.edu}}~~
and~~
Yoonbai Kim \footnote{Electronic address: {\tt yoonbai@skku.edu}}\\[4mm]
{\it Department of Physics, BK21 Physics Research Division,
and Institute of Basic Science,}\\
{\it Sungkyunkwan University, Suwon 440-746, Korea}\\
[7mm]
}

\end{center}

\vspace{5mm}

\begin{abstract}
We find exact static stringy solutions of Ho${\check {\bf r}}$ava-Lifshitz
gravity with the projectability condition but imposing the detailed balance
condition near the UV fixed point, and propose a method on constraining
the possible pattern of flows in Ho${\check {\rm r}}$ava-Lifshitz gravity
by using the obtained classical solutions.
In the obtained vacuum solutions, the parameters related to the speed of
the graviton and the coefficients of quartic spatial derivative terms
lead to intriguing effects: the change of graviton speed yields a surplus
angle and the quartic derivatives make the square of
effective electric charge negative.
The result of a few tests based on the geometries of a cone, an excess cone,
a black string, and a charged (black) string seems suggestive. For example,
the flow of constant graviton speed and variable Newton's coupling can be
favored in the vicinity of an IR fixed point, but the conclusion is indistinct
and far from definite yet. Together with the numerous classical solutions,
static or time dependent, which have already been found, the accumulated
data from various future tests will give some hints in constraining
the flow patterns more deterministic.
\end{abstract}


\newpage

\setcounter{equation}{0}
\section{Motivation and Proposal}\label{sec1}

Ho${\check {\rm r}}$ava-Lifshitz (HL) gravity of $z=3$ anisotropic scaling,
\begin{align}
t\rightarrow \ell^{z} t, \qquad x^{i}\rightarrow \ell x^{i},
\label{z3}
\end{align}
has been proposed to achieve
a theory of quantized gravity, which is believed to be
ultraviolet (UV) complete and
unitary~\cite{Horava:2009uw,Horava:2010zj,Horava:2008ih,Horava:2009if}.
It is based on the assumption that there exists renormalization group
(RG) flow
connecting a higher spatial derivative gravity such as the Lifshitz type
UV fixed point and general relativity (GR) such as the infrared (IR) fixed point.
Though nonexistence of ghost excitation is shown by analyzing the tree-level
propagator and it is believed to be power-counting renormalizable due to
sixth order spatial derivatives, the HL gravity in its various versions
encounters phenomenological inconsistencies, e.g., unwanted dynamical scalar
degree and the related strong gravitational
coupling~\cite{Horava:2009uw,Cai:2009dx,Sotiriou:2009bx},
and theoretical difficulties in
quantization procedure~\cite{Li:2009bg}. A newly proposed version of HL gravity
is intriguing since an extra local U(1) symmetry to the
foliation-preserving diffeomorphism fixes $\lambda$ to be unity and
eliminates the scalar graviton~\cite{Horava:2010zj}.

In this infant stage of HL gravity it seems difficult
to discuss specifically
the RG flows connecting the IR and UV fixed points:
Not even a single research including direct
computation of RG in the context of HL gravity has been performed yet.
It is not so puzzling since, except for some supersymmetric gauge theories
or lower-dimensional field theories,
it is difficult even in flat $(1+3)$-dimensional spacetime
to find an RG flow possessing both IR and UV fixed points.

A step aside from the issues related to propagating degrees and quantization,
there is also an intriguing research direction to find classical configurations
by solving equations of motion. In fact,
numerous classical solutions have been obtained in diverse directions,
among which the large portion consists of time-dependent cosmological
solutions~\cite{Calcagni:2009ar,Lu:2009em,Kehagias:2009is}
and static black hole
solutions~\cite{Lu:2009em,Kehagias:2009is,Cai:2009pe,Kim:2009dq,Tang:2009bu}.
Many of those are given by exact solutions despite
complication due to the higher derivative nonlinear equations.
These classical configurations contribute
obviously to the understanding classical nature of HL gravity.
However, different from the case of GR or Newtonian gravity,
direct detection of the evidences of HL gravity from the obtained classical
configurations seems extremely difficult in both
astrophysical observation in large scale and laboratory
experiments~\cite{Kim:2009dq,Dutta:2009jn,Harko:2009qr}.

A natural question is what can be the usefulness of various classical
solutions for the ultimate goal of HL gravity, a quantum gravity.
A specific question may be that whether or not the obtained
classical solutions, at least some solutions, can
be utilized to constrain possible patterns of the flows connecting
the IR and UV fixed points which are presumed to be GR and
HL gravity based on $z=3$ anisotropic scaling, respectively.
Although it is speculative, we will try to address this question in this
paper. The procedure is given as follows.
First, we consider HL gravity involving the square of the Cotton tensor
as the unique sixth order spatial derivative term, but the
coefficients of lower spatial derivative terms satisfying the
foliation-preserving diffeomorphism remain to be arbitrary.
Second, we solve the equations of motion in the coordinates compatible with
the assigned symmetries and conditions, find classical configurations,
in which some of those are given as exact solutions, and read the
corresponding geometries.
Third, we consider a multidimensional map of parameters of the theory,
and the obtained solutions are utilized in dividing the regions in the map.
There are 10 parameters composed of 8 coefficients in front of the
Lagrangian terms and 2 more constants in the classical solutions.
The selected classical solutions for application to
RG flows are required to possess the following nature:
\begin{itemize}
\item[1.] The solution is generic in the context of HL gravity.
Here ``generic'' means that the solution
is obtained not under a bizarre metric assumption, not by too many surgeries
of different geometries, and not in the presence of unphysical matters,
unphysical in the scheme of field theories with Lifshitz type anisotropic
scaling.\\[-8mm]
\item[2.] A solution is found in the wide range of parameters.\\[-8mm]
\item[3.] Solution reduces to a physically acceptable solution in the GR limit.
\end{itemize}
Note that the aforementioned static solutions fulfill the requirements.
Fourth, we examine characteristic of each solution (geometry) and classify
the corresponding region. The obtained solutions in the realm of HL gravity
are categorized as follows in comparison with GR solutions:
\begin{itemize}
\item[(I)] The usual solutions which appear as physical solutions in GR
and of which the corresponding geometries are observed
directly or/and indirectly.\\[-8mm]
\item[(II)] The solutions which are encountered in GR but of which
their astrophysical signals seem hardly or unlikely to be detected.\\[-8mm]
\item[(III)]
The solutions which are never obtained in GR or cannot be obtained in the
physical environment of GR, {\it e.g.}, unphysical solutions under
violation of the positive energy theorem.
\end{itemize}
When a parameter changes smoothly along the line of RG flow at least
in the vicinity of the IR fixed point, a classical GR solution changes to
be a classical HL solution in one of the above three categories.
If it is category (I), we may call the corresponding parameter region
an ``allowed zone.''
If it is category (II), we may call the corresponding parameter region
a ``disfavored zone.''
If it is category (III),
we may call the corresponding parameter region an ``unlikely zone.''

In the subsequent sections, we find classical solutions describing some
stringy objects and test our proposal.
In Sec.~\ref{sec2}, without matter, we solve the equations of motion
in the Painlev$\acute{\rm e}$-Gullstrand type coordinates
compatible with the projectability condition
and find exact axially symmetric solutions of which planar geometries
describe a cone with deficit angle, an excess cone with surplus angle,
and a black string with a Banados-Teitelboim-Zanelli (BTZ) type horizon.
In order to obtain the cone and excess cone configurations in GR, depicted by
the same metric, negative energy density is required in the
energy-momentum tensor side of the Einstein equations.
Therefore, the solution is unphysical in the context of GR.
In the vicinity of the IR fixed point with vanishing higher spatial derivative
terms, these solutions belong to category (III) and the parameter
region supporting these configurations may correspond to an unlikely zone.
We categorize the black string into (II) and the parameter region of this
solution may correspond to a disfavored zone. An allowed zone keeps
the speed of light propagation unity but the gravitational coupling
can be changed.
In Sec.~\ref{sec3} we solve the equations of motion in the presence of an
electrostatic field from a thin filament of charge and find a charged (black)
string solution [with Reissner-Nordstr\"{o}m (RN) type horizons].
In this metric the sum of the square of the electric charge and
a combination of coefficients in front of two curvature square terms
appears so that this square of effective charge is not positive
semidefinite.
Even in the presence of electrostatic field with positive energy density,
there is a static solution in which square of the effective charge is negative.
It means that the obtained solution is generic in HL gravity.
However, in the context of GR, the energy
density given by the square of the electric field is required to be negative
in order to get the same metric solution.
In classical electrodynamics, this can be interpreted as an attractive Coulomb
force between the same charges. Since the former violates the positive
energy theorem in GR and the latter is not allowed in the Maxwell theory,
the solution may belong to category (III) and the corresponding parameter
region may be an unlikely zone. Since the astrophysical objects with a huge
amount of excess charge can hardly be formed due to repulsive nature,
the black string
solution with the charge greater than a critical value may be categorized
into (II) and the corresponding parameter region may be a disfavored zone.
In Sec.~\ref{sec4} we summarize briefly what we obtained. Then we
discuss possible obstacles to jeopardize this speculative proposal and
the conditions to put this suggestion forward to a plausible scenario.

\setcounter{equation}{0}
\section{Vacuum Solution and Flow for Quadratic Derivative Terms}\label{sec2}

In this section we consider HL gravity with the projectability condition
but impose the detailed balance condition only in the UV limit.
In Sec.~\ref{ss21},
we look for exact classical static vacuum solutions with axial symmetry,
and the obtained configurations can describe the straight stringy objects
stretched infinitely along the $z$ axis.
In Sec.~\ref{ss22}, we discuss a possible pattern of the RG flows by
applying the obtained solutions to the method proposed in Sec.~\ref{sec1}.

The HL gravity action in (1+3) dimensions is given by
\begin{align}
S_{\rm HL}=\int dt\,d^{3}x\sqrt{g}N\left({\cal L}_{\rm IR}+{\cal L}_{\rm UV}
\right).
\label{ac2}
\end{align}
The anisotropic scaling in the UV limit \eqref{z3} lets the time specific
and reduces the diffeomorphisms
of GR to foliation-preserving diffeomorphisms on the hypersurfaces of
constant time. Thus the Arnowitt-Deser-Misner decomposition of the metric
is well equipped
\begin{align}
ds^{2}=-N^2dt^2+g_{ij}\left(dx^i+N^idt\right)\left(dx^j+N^jdt\right).
\label{ADM}
\end{align}
The IR Lagrangian density is easily determined by requiring
GR as the IR fixed point
\begin{align}
{\cal L}_{\rm IR}&=\alpha(K_{ij}K^{ij}-\lambda K^2)+\xi R+\sigma,
\label{Lir}
\end{align}
where the extrinsic curvature is given by
\begin{align}
K_{ij}\equiv
\frac{1}{2N}\left({\dot{g_{ij}}}-\nabla_iN_j-\nabla_jN_i\right),
\quad K=g^{ij}K_{ij}, \quad K^{ij}=g^{ik}g^{jl}K_{kl},\quad
{\dot{g_{ij}}}\equiv\frac{\partial g_{ij}}{\partial t}.
\end{align}
In this GR limit with general covariance,
the 4 parameters in \eqref{Lir} have the fixed values by
the Newton's constant $G$, the speed of light $c=1$, and the cosmological
constant $\Lambda$ as
\begin{align}
\lambda=1\, ,\quad \alpha=\frac{1}{16\pi Gc}\,,
\quad \xi=\frac{c}{16 \pi G}\,, \quad \sigma=-\frac{c \Lambda}{8 \pi G}\, .
\label{par}
\end{align}

Since the $z=3$ scaling \eqref{z3} still allows 5 sixth order spatial derivative
terms~\cite{Sotiriou:2009bx}, inclusion of all the terms makes
the theory less predictable. Suppose the detailed balance condition emerges
near the UV fixed point and the tensor structure of sixth order spatial
derivative terms in the action is determined by the Cotton
tensor~\cite{Horava:2009uw}
\begin{align}
C_{ij}=  \frac{\epsilon^{ikl}}{\sqrt{g}}\nabla_k\left({R^j}_l
-\frac{1}{4}R\delta^{j}_{\;l}\right).
\label{Cot}
\end{align}
The relevant deformations need not follow the detailed balance and then
the UV Lagrangian density has 4 higher spatial derivative terms
\begin{align}
{\cal L}_{\rm UV}&=\beta C_{ij}C^{ij}
+ \gamma \frac{\epsilon^{ijk}}{\sqrt{g}}R_{il}\nabla_j{R^l}_{k}
+\zeta R_{ij}R^{ij}+\eta R^2.
\label{Luv}
\end{align}
Though the physical motivation for the introduction of the detailed balance
condition is still
challenging~\cite{Cai:2009dx,Sotiriou:2009bx,Li:2009bg,Xu:2010eg},
the assignment of it restricts
the number of parameters in the Lagrangian densities in \eqref{Lir} and
\eqref{Luv} to 8 parameters given in descending order of the number of
derivatives
\begin{align}
\beta \rightarrow \gamma \rightarrow (\zeta,~\eta) \rightarrow
(\lambda,~\alpha,~\xi) \rightarrow \sigma .
\label{8pa}
\end{align}

Reflecting the difficulty in the quantization procedure~\cite{Li:2009bg},
we adopt the
projectability condition in which the lapse function $N$ depends only
on the time coordinate $t$. Then the time reparametrization, a symmetry
transformation in the projectable version of HL gravity, fixes the lapse
function to be unity, $N=1$. Since we are interested in stringy static
objects,
we assume rotational symmetry in the $(r,\theta)$ plane and take the
Painlev$\acute{\rm e}$-Gullstrand type coordinates of which the metric is
compatible with the projectability condition
\begin{align}
ds^2=-dt^2+e^{-2f(r)}\left[dr+n(r)dt\right]^2+r^2d\theta^2+e^{2h(r)}dz^2.
\label{met5}
\end{align}
Inserting this metric into the HL action \eqref{ac2}, we have
\begin{align}
S_{\rm HL}=&2\pi\int dt\, dr dz\, r\, e^{h-f}\Bigg\{\alpha(1-\lambda)\left(
n'^{2}-2nn'f'+n^2f'^2+n^2h'^2+\frac{n^2}{r^2}\right)\nonumber\\
&\hspace{0mm}-2\alpha \lambda \left[\left(nn'-n^2f'\right)\left(\frac{1}{r}+h'\right)
+\frac{n^2h'}{r}\right]\nonumber\\
&\hspace{0mm}+\frac{\beta}{2}e^{4f}\left[\frac{h'-f'}{r^2}+\frac{f''-h''+2(f'^2-h'^2)}{r}
-2f'^2h'-3f'h''-2f'h'^2-f''h'-2h'h''-h'''\right]^2\nonumber\\
&\hspace{0mm}+\zeta e^{4f}\left[\left(\frac{f'}{r}+f'h'+h''+h'^2\right)^2+\left(\frac{f'}{r}+\frac{h'}{r}\right)^2
+\left(\frac{h'}{r}+f'h'+h''+h'^2\right)^2\right]\nonumber\\
&+4\eta e^{4f}\left(\frac{f'}{r}+\frac{h'}{r}+f'h'+h''+h'^2\right)^2-2\xi e^{2f}\left(\frac{f'}{r}+\frac{h'}{r}+f'h'+h''+h'^2\right) +\sigma\Bigg\},\label{ac4}
\end{align}
where the prime $'$ denotes $d/dr$.
Note that the parity-violating fifth order derivative term in \eqref{Luv}
vanishes, and then
the action \eqref{ac4} involves 7 parameters except $\gamma$.
For later convenience we introduce
the action of matter fields,
\begin{align}
S_{\rm M}=&2\pi\int dt\,dr dz \,r\, e^{h-f}\,{\cal L}_{\rm M}(n,f,h).
\label{mac3}
\end{align}

The equations of motion are read by varying the actions
\eqref{ac4}--\eqref{mac3} with respect to the metric functions.
The variation of $n$ gives
\begin{align}
&(1-\lambda)\left[1+r^2f''-\frac{rn'}{n}\left(1-rf'+rh'\right)-\frac{r^2n''}{n}\right]
+\lambda r^2h''+rf'(1+rh')+r^2h'^2\nonumber\\
&=-\frac{r^2}{2\alpha n}\frac{\partial {\cal L}_{\rm M}}{\partial n},
\label{neq2}
\end{align}
but we omit the equations of $f$ and $h$, which are too lengthy.

In order to obtain exact solutions under minimal restriction, we fix a
parameter among the remaining 7 parameters.
Specifically, reflecting the revival
of general covariance in the IR limit, we choose
\begin{align}
\lambda=1.
\label{l1}
\end{align}
In HL gravity possessing an extra U(1) gauge symmetry~\cite{Horava:2010zj},
the fixation of \eqref{l1} is naturally forced.
If some exact solutions are obtained under this choice,
those are valid in the IR regime. Under \eqref{l1},
the Eq. \eqref{neq2} is simplified
\begin{align}
f'=-\frac{r}{1+rh'}\left(h''+h'^2+\frac{1}{2\alpha n}
\frac{\partial {\cal L}_{\rm M}}{\partial n}\right),
\label{neq3}
\end{align}
however the other equations of $f$ and $h$ are still left
in complicated forms
\begin{align}
&-2 r^5 \sigma -4 r^4 \alpha  n n' \left(1+r h'\right)-4r^4 \alpha  n^2 \left( h'+rh'^2+rh''\right)-4 e^{4 f} r^4 \beta  h'^5\nonumber\\
&+ e^{4 f} \Bigg\{2 r^2 \beta  f'^3 (1-r h') \left[r^2 \left(2 h'^2+h''\right)
-1\right]-2 r^3 h'^4 \left[r^2 \beta
\left(f''+2 h''\right)-5 \beta -2 r^2 (\zeta +2 \eta )\right]
\nonumber\\
&+2 r^2 h'^3 \left[-3 \beta -2 r^2 \zeta +r^2 \beta
\left(4 f''+h''+2 r f'''-r h'''\right)\right]- r f'^2
\Bigg[-5 \beta +12 r^2 (\zeta +2 \eta )\nonumber\\
&+r \Big[8 r^3 \beta  h'^4-12 r^2 \beta  h'^3+2 r h'^2 \left(13 \beta +6 r^2 (\zeta +2 \eta )+r^2 \beta  \left(-9 f''+2 h''\right)\right)+r \beta  \big[-18 f''\nonumber\\
&+h'' \left(20+19 r^2 h''\right)+14 r h^{'''}\big]+2 h' \big(\beta +6 r^2 (\zeta +4 \eta )+r^2 \beta  \left(18 f''+7 h''-7 r h'''\right)\big)\Big]\Bigg]\nonumber\\
&+ r h'^2 \Bigg[-7 \beta +8 r^2 (\zeta +\eta )+r^2 \bigg[9 r^2 \beta  f''^2+4 f'' \Big(-2 \left(2 \beta +r^2 (\zeta +2 \eta )\right)+3 r^2 \beta  h''\Big)\nonumber\\
&-2 \Big(\left(9 \beta +4 r^2 (\zeta +2 \eta )\right) h''+2 r^2 \beta  h''^2+r \beta  \big(2 f'''-5 h'''-r (f''''+2 h'''')\big)\Big)\bigg]\Bigg]\nonumber\\
&-2  f' \Bigg[3\beta -4 r^2 (\zeta +2 \eta )(1-r^3h'^3)+r \bigg[2 r^4 \beta  h'^5-2 r^3 \beta  h'^4+r^2 h'^3 \left(9 \beta-r^2 \beta  (4 f''-7 h'')\right)\nonumber\\
&+r h'^2 \Big(-9 \beta +8 r^2 (\zeta +3 \eta )-r^2 \beta  \left(-4 f''+7 h''+6 r (f'''+h''')\right)\Big)+r \Big(12 r^2 \beta  h''^2\nonumber\\
&+\beta  f'' \left(7+11 r^2 h''\right)+h'' \left(4 r^2 (\zeta +4 \eta )-7 \beta +7 r^3 \beta  h'''\right)+r \beta  \left(7 h'''-6 f'''+5 r h''''\right)\Big)\nonumber\\
&+h' \Big(-\beta +4 r^2 (\zeta +2 \eta )+r^2\left(25 \beta +8 r^2 (\zeta +2 \eta )\right) h''+4 r^4 \beta  h''^2-r^2\beta  f'' \left(7+11 r^2 h''\right)\nonumber\\
&+r^3 \beta  \big(12 f'''+6 h'''-5 r h''''\big)\Big)\bigg]\Bigg]-r^3 \left(9 \beta -4 r^2 (\zeta +2 \eta )\right) h''^2-9 r^3 \beta  f''^2+4 r^5 \beta  h''^3\nonumber\\
&+2rf'' \Big(-3 \beta +4 r^2 (\zeta +2 \eta )+r^2 \beta  \left(6 h''+4 r^2 h''^2+7 r h'''\right)\Big)\nonumber\\
&+2rh'' \Big(3 \beta +2 r^2 (\zeta +4 \eta )+r^3 \beta  \left(4 f'''+8 h'''+r h''''\right)\Big)\nonumber\\
&+r^2 \bigg[\left(4 r^2 (\zeta +4 \eta )-6 \beta \right) h'''+4 \beta  f'''+r^3 \beta  h'''^2-2 r \beta  \left(f''''-2 h''''-r h'''''\right)\bigg]\Bigg\}\nonumber\\
&+2 e^{2 f} h' \Bigg\{2 r^4 \xi +e^{2 f} \Bigg[3 \beta +2 r^2 (\zeta +4 \eta )+r^2 \Big[-9 r^2 \beta  f''^2-2 r^2 \beta  h''^2+r^3 \beta  \left(2 f'''+h'''\right)\nonumber\\
&+2 h'' \Big(4 \left(\beta -r^2 (\zeta +3 \eta )\right)\Big)+f''\Big(-2 \left(\beta +2 r^2 (\zeta +4 \eta )\right)+r^2 \beta  \left(-8 h''+7 r h'''\right)\Big)\nonumber\\
&+r \Big(2 \beta  f'''-\left(11 \beta +4 r^2 (\zeta +2 \eta )\right) h'''-r \beta  \left(2 f''''+h''''-r h'''''\right)\Big)\Big]\Bigg]\Bigg\}\nonumber\\
&= 2r^5\left({\cal L}_{\rm M}-\frac{\partial {\cal L}_{\rm M}}{\partial f}\right) ,
\label{feq3}
\end{align}
\begin{align}
&2 r^5 \left(\sigma +2 \alpha  n'^2\right)+2 r^4 \alpha  n^2 \left(-2 f'+2 r f'^2-2 r f''\right)+4 r^4 \alpha  n \big(\left(2-3 r f'\right) n'+r n''\big)\nonumber\\
&-2 e^{4 f} r^3 h'^4 \Big(5 \beta +2 r^2 (\zeta +2 \eta )-r^2 \beta  \left(3 f''+2 h''\right)\Big)-4 e^{4 f} r^3 \beta  f'^4 \Big(11+20 r^2 h''+rh' \left(10 r h'-1\right)\Big)\nonumber\\
&-2 e^{4 f} r^2 h'^3 \Big( 2 r^2 (\zeta +4 \eta )-8 \beta+r^2 \beta  \big(13 f''-7 h''+r \left(f'''-3 h'''\right)\big)\Big)+4 e^{4 f} r^4 \beta  h'^5\nonumber\\
&+24 e^{4 f} r^4 \beta  f'^5 \left(1-r h'\right)+2 e^{4 f} r^2 f'^3 \Bigg\{12 \left(2 \beta +r^2 (\zeta +4 \eta )\right)-r \Big[44 r \beta  h'^2+2 r^2 \beta  h'^3\nonumber\\
&+h' \Big(r^2 \beta  \left(64 f''+81 h''\right)-3 \left(7 \beta +8 r^2 (\zeta +2 \eta )\right)\Big)+r \beta  \left(-64 f''+22 h''+51 r h'''\right)\Big]\Bigg\}\nonumber\\
&+e^{4 f} r f'^2 \Bigg\{-41 \beta -8 r^2 (\zeta +5 \eta )+r \Big[-64 r^2 \beta  h'^3+16 r^3 \beta  h'^4+2 r h'^2 \Big(35 \beta +22 r^2 (\zeta +2 \eta )\nonumber\\
&-r^2 \beta  \left(59 f''+14 h''\right)\Big)-2 h' \Big(28 \beta -r^2 (26 \zeta +64 \eta )+r^2 \beta  \left(7 f''+127 h''+37 r f'''+65 r h'''\right)\Big)\nonumber\\
&+r \Big(\left(68 \beta +88 r^2 (\zeta +2 \eta )\right) h''-22 \beta  f'' \left(5+11 r^2 h''\right)- r \beta  \left(70 h'''+62 r h''''-74 f'''\right)\nonumber\\
&-111 r^2 \beta  h''^2\Big)\Big]\Bigg\}-e^{4 f} r h'^2 \Bigg\{r^2 \Big[27 r^2 \beta  f''^2+2 f'' \left(- 12 \beta -8 r^2 (\zeta +2 \eta )+7 r^2 \beta  h''\right)\nonumber\\
&+2 \Big(\left(17 \beta +4 r^2 (\zeta +2 \eta )\right) h''-6 r^2 \beta  h''^2+r \beta  \big(11 f'''+13 h'''+r \left(3 f''''+h''''\right)\big)\Big)\Big]\nonumber\\
&+5 \beta +4 r^2 (\zeta +2 \eta )\Bigg\}-2 e^{4 f} h' \Bigg\{r^2 \Big[h'' \Big(2 \left(\beta -r^2 (5 \zeta +4 \eta )\right)+r^3 \beta  \left(20 f'''+7 h'''\right)\Big)\nonumber\\
&+4 r^2 \beta  f''^2+21 r^2 \beta  h''^2+f'' \Big(2 \left(5 \beta -r^2 (5 \zeta +12 \eta )\right)+r^2 \beta  \left(58 h''+17 r f'''+29 r h'''\right)\Big)\nonumber\\
&+r^2 \beta  \Big(f''''+13 h''''+r \left(f'''''+3 h'''''\right)\Big)-r\left(3 \beta+ 4 r^2 (\zeta +2 \eta )\right)\left( f'''-2h'''\right) \Big]\nonumber\\
&+9 \beta -4 r^2 (\zeta +2 \eta )\Bigg\}+2 e^{2 f} f' \Bigg\{e^{2 f} \left(9 \beta +2 r^2 (\zeta +4 \eta )\right)-2 r^4 \xi +e^{2 f} r \Bigg[4 r^3 \beta  h'^4+2 r^4 \beta  h'^5\nonumber\\
&-r^2 h'^3 \Big(13 \beta +4 r^2 (\zeta +2 \eta )+r^2 \beta  \left(4 f''-17 h''\right)\Big)-r h'^2 \Big(8 \beta -4 r^2 (3 \zeta +2 \eta )\nonumber\\
&+r^2 \beta  \left(62 f''+64 h''+22 r f'''+7 r h'''\right)\Big)-h' \bigg[-23 \beta +12 r^2 (\zeta +2 \eta )+r^2 \Big[47 r^2 \beta  f''^2\nonumber\\
&-3 \left(13 \beta +12 r^2 (\zeta +2 \eta )\right) h''+15 r^2 \beta  h''^2+f'' \left(-23 \beta -28 r^2 (\zeta +2 \eta )+113 r^2 \beta  h''\right)\nonumber\\
&+r \beta  \left(6 f'''+70 h'''+10 r f''''+23 r h''''\right)\Big]\bigg]+r \bigg[-57 r^2 \beta  h''^2+h'' \Big(-25 \beta +2 r^2 (15 \zeta +32 \eta )\nonumber\\
&-5 r^3 \beta  \left(9 f'''+11 h'''\right)\Big)+f'' \Big(31 \beta +14 r^2 (\zeta +4 \eta )-r^2 \beta  \left(45 h''+73 r h'''\right)\Big)+47 r^2 \beta  f''^2\nonumber\\
&+r \Big(-17 \beta f'''+6 \left(3 \beta +4 r^2 (\zeta +2 \eta )\right) h'''+r \beta  \left(10 f''''-9 \left(2 h''''+r h'''''\right)\right)\Big)\bigg]\Bigg]\Bigg\}\nonumber
\end{align}
\begin{align}
&-e^{4 f} r \Bigg\{2 h'' \Big(4 r^2 (\zeta +2 \eta )-9 \beta +r^3 \beta  \left(10 f'''+29 h'''+6 r f''''+9 r h''''\right)\Big)\nonumber\\
&-3 r^2  h''^2\left(3 \beta +4 r^2 (\zeta +2 \eta )\right)+4 r^4 \beta  h''^3+7 r^2 \beta  f''^2 \left(3+8 r^2 h''\right)+2 f'' \Big[9 \beta +2 r^2 (\zeta +4 \eta )\nonumber\\
&+r^2 \Big(-4 \left(3 \beta +4 r^2 (\zeta +2 \eta )\right) h''+24 r^2 \beta  h''^2+r \beta  \left(-17 f'''+19 h'''+14 r h''''\right)\Big)\Big]\nonumber\\
&+r \Big[4 \left(3 \beta -4 r^2 (\zeta +2 \eta )\right) h'''+13 r^3 \beta  h'''^2+2 f''' \left(-5 \beta -2 r^2 (\zeta +4 \eta )+13 r^3 \beta  h'''\right)\nonumber\\
&+2 r \Big(2 \beta  f''''-\left(3 \beta +4 r^2 (\zeta +2 \eta )\right) h''''-r \beta  \left(f'''''-3 h'''''-r h''''''\right)\Big)\Big]\Bigg\}\nonumber\\
&=-2r^5\left({\cal L}_{\rm M}
+\frac{\partial {\cal L}_{\rm M}}{\partial h}\right).
\label{heq3}
\end{align}
The above equations are derived by variation of the action
\eqref{ac4}, but are checked to be the same as the equations of motion
obtained from \eqref{ac2} after inserting \eqref{ADM} under the projectability
condition since the metric \eqref{met5} possesses good symmetries.

\subsection{Exact solution}\label{ss21}

From here on, we focus on the case of $\lambda=1$ and obtain some
stringy configurations as exact solutions in the absence of matter field,
${\cal L}_{\rm M}=0$.
Suppose we have a constant solution of $f$ as
\begin{align}
f=(\ln f_0)/2,\qquad (f_{0}>0).
\label{fcon}
\end{align}
Then \eqref{neq3} becomes an equation of $h(r)$ and we obtain
\begin{align}
h(r)=\ln[h_0(r-r_0)],
\label{hsol2}
\end{align}
where $h_{0}$ and $r_{0}$ are two integration constants,
and $h_0$ can be set to be unity by a rescaling of the $z$ coordinate.
Plugging \eqref{hsol2} and the constant $f$ solution \eqref{fcon}
into the remaining two equations
\eqref{feq3}--\eqref{heq3}, we find the exact solution for the vanishing
integration constant, $r_{0}=0$,
\begin{align}
h(r)=&\ln r,
\label{hsol}\\
n(r)=&\pm \sqrt{-\frac{\sigma}{6\alpha}r^2
+f_0\frac{\xi}{\alpha}+\frac{n_0}{r}-\frac{f_0^2(5\zeta+14\eta)}{\alpha r^2}}\, ,
\label{snf}
\end{align}
where $n_0$
is another integration constant.
Note that the Cotton tensor \eqref{Cot} vanishes for the obtained solution
and it is the reason why
the exact solution in \eqref{fcon}, \eqref{hsol}, and \eqref{snf}
does not involve $\beta$ dependence.
Finally we obtain the following metric from \eqref{met5}
\begin{align}
ds^2=-dt^2+\frac{1}{f_0}\left(dr\pm\sqrt{-\frac{\sigma}{6\alpha}r^2
+f_0\frac{\xi}{\alpha}+\frac{n_0}{r}-\frac{f_0^2(5\zeta+14\eta)}{\alpha r^2}}\,\,dt\right)^2+r^2\left(d\theta^2+dz^2\right),
\label{mev}
\end{align}
where the metric involves a constant metric, $f_{0}$, and
an integration constant, $n_{0}$.
We will make a few comments on the solution \eqref{mev} and then figure out
the physical meaning of those in what follows.

If we regard the $z$ coordinate as a compact one, {\it e.g.}, an $S^{1}$,
the $(\theta , z)$ surface forms a $T^{2}$. In the nonprojectable version of
HL gravity involving the lapse function with spatial coordinate dependence,
there is another equation for $N(r)$. Then \eqref{mev} need not always be
a solution but, actually, is a solution only for the restricted parameters,
$3\zeta +8\eta=0$. Since the nonprojectable version is a different theory
from the projectable version, we shall focus on the solution \eqref{mev}
of the projectable HL gravity in this paper.

To compare, under a static metric in the Poincar\'{e} coordinates
$ds^{2}=-B(r)e^{2\delta(r)}d{\tilde t}^{2}
+[dr^{2}/B(r)]+r^{2}(d\theta^{2}+dz^{2})$, we directly solve the Einstein
equations without the matter fields and obtain
\begin{align}
ds^{2}=-\left(-\frac{\Lambda}{3}r^{2}-\frac{GM}{r}\right)
d{\tilde t}^{2}+\frac{dr^{2}}{-\frac{\Lambda}{3}r^{2}
-\frac{GM}{r}}+r^{2}(d\theta^{2}+dz^{2}),
\label{Em2}
\end{align}
where $\Lambda$ is a cosmological constant and $M$ is an integration constant.
Understanding the physical meaning of the 5 parameters in the solution
\eqref{mev}, $\sigma, \xi, \alpha, f_0,$ and $n_0,$ we perform
the coordinate transformation,
\begin{align}
d\tilde{t}=\frac{1}{\sqrt{F_0}}\left(dt\mp\frac{\sqrt{F_0-B}}{B}dr\right),
\qquad (F_{0}>0),
\label{tr1}
\end{align}
and the resultant metric is
\begin{align}
ds^2=-dt^2+\frac{1}{F_0}\left(dr\pm\sqrt{\frac{\Lambda}{3}r^2
+F_0+\frac{GM}{r}}\,\,dt\right)^2
+r^2\left(d\theta^2+dz^2\right).
\label{Em1}
\end{align}
Let us take the IR limit of vanishing higher derivative terms
in \eqref{mev}, $\zeta=\eta=0$, and compare it with \eqref{Em1}.

Comparing the first terms proportional to $r^{2}$ in \eqref{mev} and
\eqref{Em1},
the identification in the GR limit, $\sigma/6\alpha=-\Lambda/3$, matches
the relations in \eqref{par} and then
the first term is interpreted as a cosmological constant term.
There is an $r^{2}$ factor in front of $dz^{2}$,
and this warp factor seems unavoidable for straight stringy solutions
in $(1+3)$-dimensional HL
gravity~\cite{Cho:2009fc}. In addition it makes the integration constant term
inversely proportional to the radial coordinate, $-n_0/r$, and this $1/r$ behavior
is formally nothing but the mass term in the $(1+3)$-dimensional
Schwarzschild solution with spherical symmetry.
The metric function in \eqref{Em2} does not have a constant term and
thus, in anti-de Sitter (AdS) spacetime, it naturally possesses a horizon at
$r_{{\rm H}}=(-3GM/\Lambda)^{1/3}\sim (6\alpha n_0/\sigma)^{1/3}$
for positive mass $M>0$ similar to the case of the BTZ black hole~\cite{Banados:1992wn}.
Therefore, in the GR limit, the integration constant of the $1/r$ term
in \eqref{Em1} is identified with the usual mass parameter in \eqref{Em2}
as $n_{0}=GM$. It can also be confirmed by
computing the Kretschmann invariant for \eqref{Em2}, which
shows a physical singularity at the origin for nonvanishing $GM$
\begin{align}
R^{\mu\nu\rho\sigma}R_{\mu\nu\rho\sigma}
= \frac{12G^2M^2}{r^6}+\frac{8\Lambda^2}{3}.
\label{4Kr}
\end{align}
Though there looks to be a constant piece $F_{0}$ in \eqref{Em1}, it is an
artifact of the coordinate transformation \eqref{tr1} in GR and thus it
does not appear as the original metric in \eqref{Em2}.
In HL gravity, on the other hand, the transformation which mixes the $t$ and
$r$ coordinates is not allowed as a symmetry transformation. Thus,
$f_{0}$ in the HL metric \eqref{mev} should be treated as another integration
constant.

From here on we consider an IR regime where the 2 parameters,
$\alpha$, $\xi$, are slightly away from GR but higher spatial
derivative
terms are left to vanish, $\zeta=\eta=0$. Since we only have an exact solution
for $\lambda=1$, we keep $\lambda$ to be unity for tractability.
In the metric \eqref{Em1} a small variation of the coefficients in
the $r^{2}$ and $1/r$ terms induces only slightly and
qualitatively the same change in \eqref{Em2}.
On the other hand, a tiny deviation of $\xi/\alpha$ from unity changes
the situation drastically and qualitatively
provided the integration constant is not
fine-tuned to an extremely small value, $f_{0}\nrightarrow 0^{+}$.

\begin{figure*}[ht]\centering
\vspace{10mm}
\scalebox{1.2}[1.2]{
\includegraphics[width=50mm]{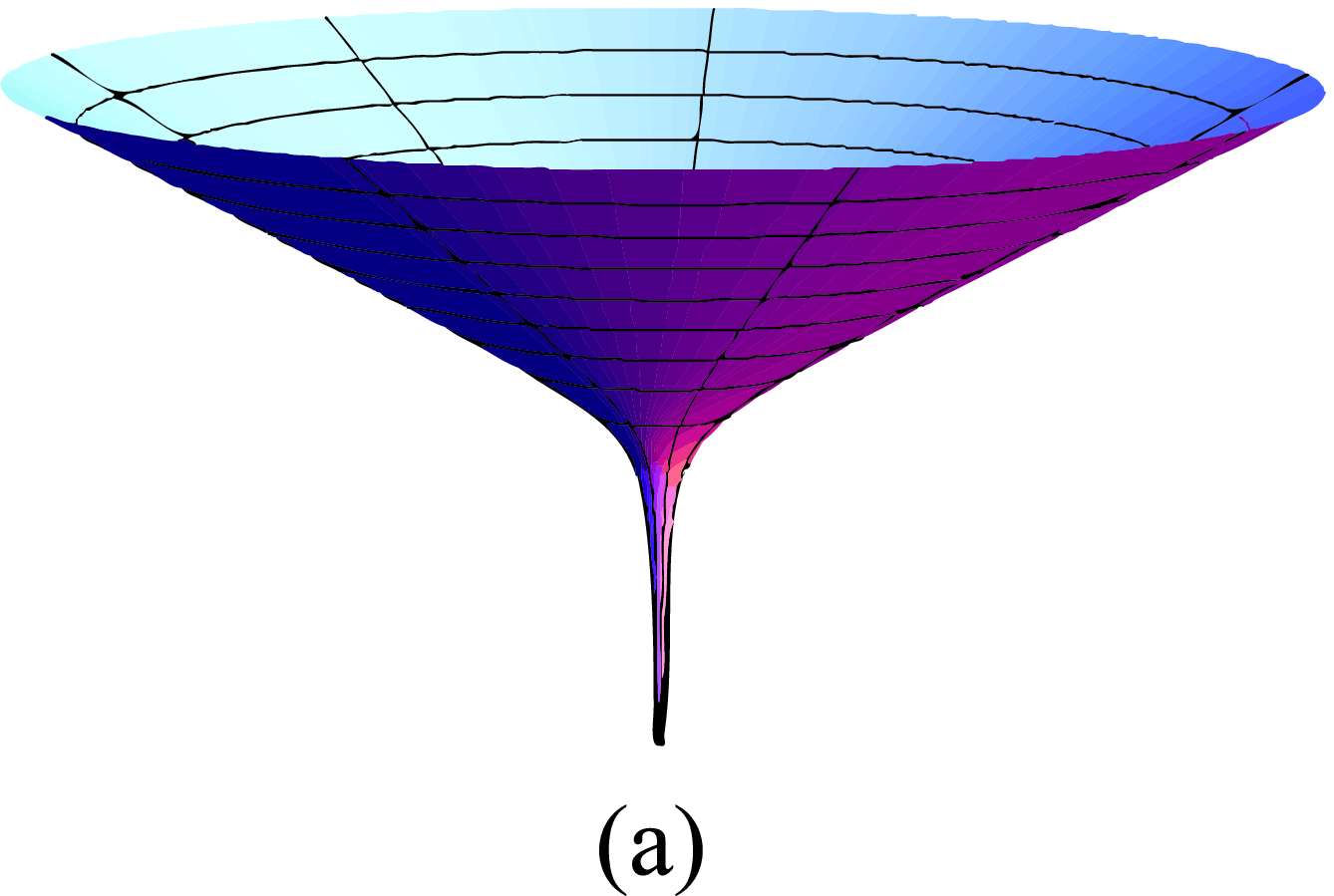}\hspace{20mm}\includegraphics[width=50mm]{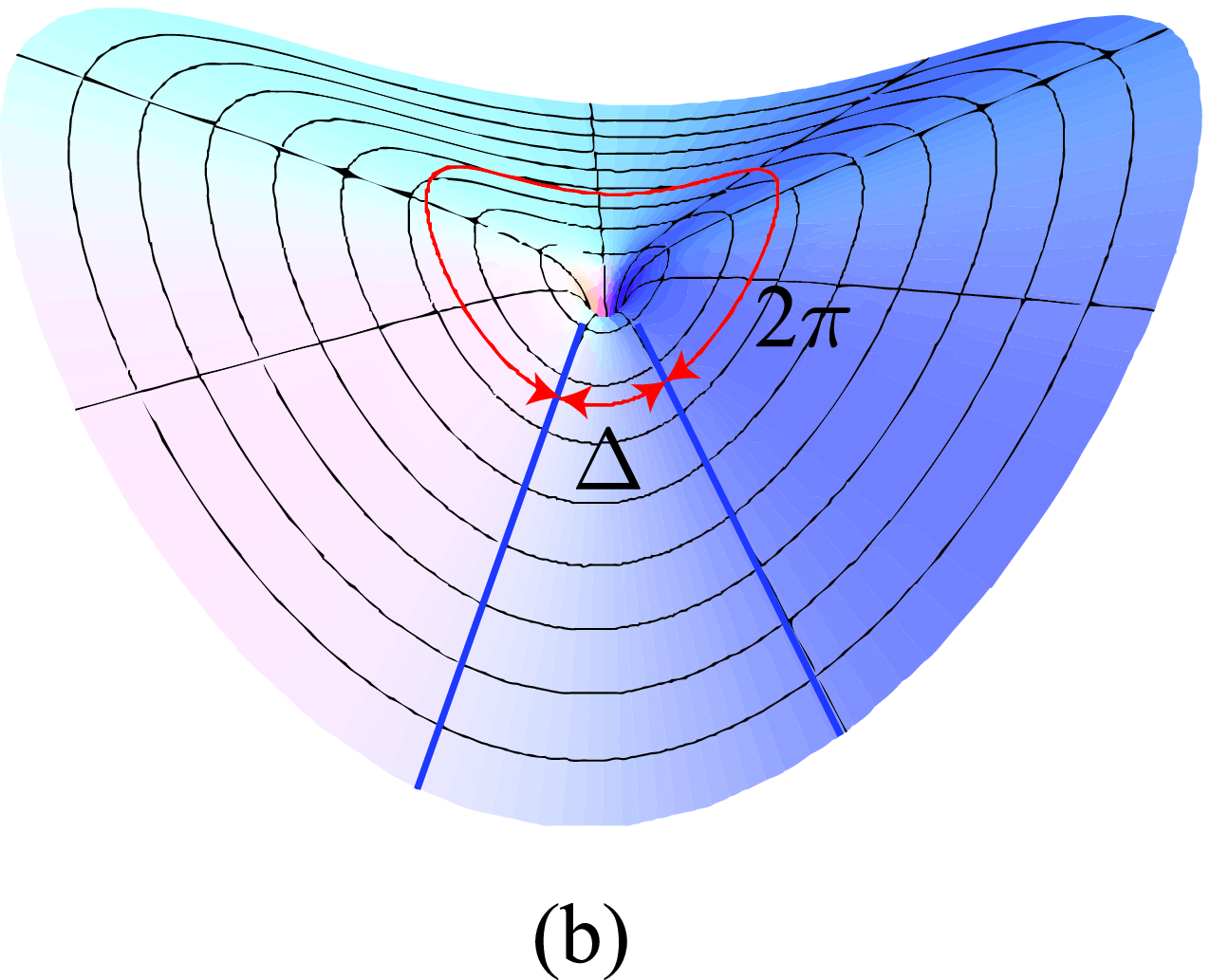}
}\\
\caption{\small Surfaces of the $(r,\theta)$ coordinates for vanishing
cosmological constant $\sigma=0$, involving a singularity at the origin $r=0$
for $n_{0}>0$.
(a) An asymptotic cone of deficit angle
$\Delta=2\pi(1-\sqrt{f_0\delta}\,)=\pi/4$ for $0<f_{0}\delta<1$ and $n_{0}=3$.
(b) An asymptotic excess cone of surplus angle
$\Delta=2\pi(\sqrt{f_0\delta}-1)=\pi/4$ for $f_{0}\delta>1$ and $n_{0}=3$.
}
\label{fig01}
\end{figure*}
To illustrate the geometry of this case, let us introduce a parameter
$\delta=1-\xi/\alpha$ measuring variation of the graviton speed from unity.
First, we consider the parameter region of positive $\delta$
($\xi/\alpha<1$), where the graviton speed becomes smaller than
unity even in a flat spacetime limit.
To make the discussion simpler without spoiling physics we can
take the limit of zero cosmological constant $\sigma=0$ and choose zero
integration constant $n_{0}=0$ (zero ``mass'').
Performing the coordinate transformation which preserves
the foliation-preserving diffeomorphism and is consistent with the
projectability condition,
\begin{align}
dt=\sqrt{f_{0}}\, d{\check t}\pm\frac{\sqrt{\frac{\xi}{\alpha}}\, dr}{
\sqrt{f_{0}}\left(1-\frac{\xi}{\alpha}\right)},
\end{align}
we obtain the metric which is rescaled to be
\begin{align}
ds^{2}=& -f_{0}\delta\, d{\check t}^{2}+\frac{dr^{2}}{f_{0}\delta}
+r^{2}(d\theta^{2}+dz^{2})
\label{Em22}\\
=&- d{\bar t}^{2}+d{\bar r}^{2}+{\bar r}^{2}(d{\bar \theta}^{2}+d{\bar z}^{2})
,
\label{Em3}
\end{align}
where the coordinates in the first and second lines are related by
\begin{align}
d{\check t}=\frac{d{\bar t}}{\sqrt{f_0\delta}},\quad dr=\sqrt{f_0\delta}\,
d{\bar r},\quad d\theta=\frac{d{\bar \theta}}{\sqrt{f_0\delta}}\,~~(0\le {\bar \theta}<2\pi\sqrt{f_0\delta}\,),
\quad dz=\frac{d{\bar z}}{\sqrt{f_0\delta}}.
\end{align}
For $f_{0}$ satisfying
$0<f_{0}\delta<1$, the $({\bar r},{\bar \theta})$-space is conic with
deficit angle $\Delta=2\pi(1-\sqrt{f_0\delta}\,)$. For $f_{0}\delta$
greater than unity, it becomes
an excess cone with surplus angle
$\Delta=2\pi(\sqrt{f_0\delta}-1)$.
This can also be confirmed by analyzing the geodesic equation of a test body
from the metric \eqref{mev}. The orbit equation for $u\equiv 1/r$
corresponding to \eqref{mev} in the limit of $\zeta=\eta=\sigma=n_{0}=0$ is
\begin{align}
\frac{d^{2}u}{d\theta^{2}}+(f_{0}\delta)u=0,
\end{align}
and the orbit,
\begin{align}
u(\theta)=u_{0}\cos\left(\sqrt{f_{0}\delta}\,\theta\right),
\end{align}
is closed not at $2\pi$ but at
$2\pi\sqrt{f_{0}\delta}\,$.
It exactly coincides with the deficit (surplus) angle read from the
metric \eqref{Em3}.

If we turn on nonvanishing positive $n_{0}$, there is the aforementioned
singularity at the origin but the $(r, \theta)$ plane approaches a cone
(an excess cone) with deficit (surplus) angle in its asymptote irrespective
of the value of $n_{0}$.
See Fig.~\ref{fig01} from \eqref{mev}
with $\sigma=0$, $f_{0}\delta>0$, and $n_{0}>0$.
In the limit of $f_{0}\delta\rightarrow 1$ and $n_{0}\rightarrow 0$,
both the deficit and surplus
angles approach zero and the geometry of $(r,\theta)$ coordinates is nothing
but a flat plane.
If the $z$ coordinate is compact like an $S^{1}$, both the $(r,\theta)$ and
$(r,z)$ planes form cones with a deficit (surplus) angle.

Astrophysical implication of deficit or surplus angle from a straight stringy
object is simple and can easily
be detected if the size of the angle is much larger than the known astronomical
bound which is about 200 $\mu$arc\,s ($\sim 10^{-9}$ rad)~\cite{CHA}.
When a conic geometry with deficit angle is formed, the light from a star
behind the singular stringy source at the apex propagates straight and arrives
at a static observer who detects double images projected behind the
source~\cite{Vilenkin:1981zs}.
If a cosmic string traverses a star behind it, the light curve increases
up by a factor of 2 due to microlensing by the string during the period of
traversing~\cite{Kuijken:2007ma}.
In the case of an excess conic geometry with surplus angle, under the same
situation as the deficit angle, the static observer tracking down the
trajectory of the star experiences a sudden disappearance of its image
for a while and reappearance at a distant point over the source of the surplus
angle~\cite{Kim:2009dq}. The geometry we are dealing with involves
an undetermined constant solution $f_{0}$ \eqref{fcon}, for which
there is no way to determine its
value or range at the moment. We can only say that its effect may be
detected for some $f_{0}$ sufficiently larger than $1/\delta$, which must be
very large near GR. Though there does not seem to be any report on the astronomical
observation of the surplus angle in the physics world governed by GR and its
Newtonian limit, we are actually surrounded by surplus angles on the leaves
in our daily life or in the world of plants [see Fig.~\ref{fig01}-(b)
of which the $n_{0}\rightarrow 0$ limit can be viewed as a leaf of locus].

Second, we consider the parameter region of negative $\delta$
($\xi/\alpha>1$), where the graviton speed becomes larger than
unity even in the flat spacetime limit. To keep the correct spacetime signature
at asymptotic space, we choose anti-de Sitter spacetime of a negative
cosmological constant $\sigma>0~(\Lambda<0)$ but again turn off the mass
$n_0=0~(GM=0)$ for simplicity.
Then, in a $(1+2)$-dimensional spacetime of $dz=0$, the metric,
\begin{align}
ds^{2}=-dt^{2}+\frac{1}{f_{0}}\left(dr\pm\sqrt{-\frac{\sigma}{6\alpha}r^{2}
+f_{0}\frac{\xi}{\alpha}}\, dt\right)^{2}+r^{2}d\theta^{2},
\label{bsm}
\end{align}
describes a BTZ black hole with a horizon at
$r_{{\rm H}}=\sqrt{6\alpha f_{0}|\delta|/\sigma}\,,$ which moves to zero
in the limit of GR. In $(1+3)$ dimensions, \eqref{bsm} with $r^{2}dz^{2}$
describes a straight black string.

In the context of GR the planar geometry of a cone (an excess cone) is
generated by a $\delta$-function source of positive (negative) energy density.
In three spatial dimensions the conic geometry is obtained
by assuming a thin infinitely stretched source, and it is applied for
depicting cosmic strings in the early universe~\cite{Vilenkin:1981zs}.
In HL gravity we find a solution of constant $f$ \eqref{fcon} and
logarithmic $h$ \eqref{hsol} under a physically allowable parameter
\eqref{l1}. Then, for $\zeta=\eta=0$ and $\sigma=0$, the only equation
for the shift function $n(r)/f_{0}$ reduces to
\begin{align}
(rn^{2})^{'}=f_{0}\frac{\xi}{\alpha},
\label{veq}
\end{align}
and it supports the conic (excess conic) geometry with deficit (surplus)
angle in \eqref{Em3} as a static vacuum solution.
Let us try to obtain the same solution in the context of GR. We assume
the metric with undetermined shift function,
\begin{align}
ds^{2}=-dt^{2}+\frac{1}{f_{0}}\left[dr+n(r)dt\right]^{2}+r^{2}(d\theta^{2}
+dz^{2}),
\label{mn}
\end{align}
and then the Einstein equation,
$G^{\mu}_{\;\;\nu}=8\pi GT^{\mu}_{\;\;\nu}$,
becomes
\begin{align}
(rn^{2})^{'}=f_{0}-8\pi Gr^{2}T^{t}_{\; t}.
\label{Ett}
\end{align}
In order to obtain a cone (an excess cone) in the $(r,\theta)$ plane
orthogonal to the string with a warp factor along the string direction,
$r^{2}dz^{2}$ in GR, we should add the negative energy density
\begin{align}
8\pi G(-T^{t}_{\;\; t})=-f_{0}\delta/r^{2}<0.
\label{-en}
\end{align}
Amazingly only the negative energy
density of $f_{0}\delta >0$ can generate either the solution of the deficit
angle $(0<f_{0}\delta<1)$ or that of the surplus angle $(f_{0}\delta>1)$ in GR,
which violates the positive energy theorem. In HL gravity, on the other hand,
we assumed the absence of matter, ${\cal L}_{{\rm M}}=0$, and thus the obtained
solution is generic.

To read the singularity in HL gravity, we look into
the curvature tensor $R_{ijkl}$ and the extrinsic curvature $K_{ij}$.
Therefore possible candidates can be two scalar quantities invariant
under foliation-preserving diffeomorphism, which are
a three-dimensional analog of the Kretschmann invariant,
$R^{ijkl}R_{ijkl}=4R_{ij}R^{ij}-R^2$,
and a square of the extrinsic curvature, $K_{ij}K^{ij}$,
which is a kinetic part of the $(1+3)$-dimensional scalar curvature.
Substitution of the metric \eqref{met5} and the vacuum solution
\eqref{mev} leads to
\begin{align}
R^{ijkl}R_{ijkl}
=&4e^{4f}\left[\frac{f'^2+h'^2}{r^2}+\left(f'h'+h'^2+h''\right)^2\right]
\label{Kre}\\
=&\frac{4f_0^2}{r^4},
\label{Krev}\\
K_{ij}K^{ij}=&n^2\left[\frac{1}{r^2}+\left(f'-\frac{n'}{n}\right)^2+h'^2\right]\\
&\hspace{-15mm}=\frac{1}{r^2}\left\{-\frac{\sigma}{3\alpha}r^2+\frac{2f_0\xi }{\alpha}+\frac{2n_0}{r}
-\frac{2f_0^2(5\zeta+14\eta)}{\alpha r^2}+\frac{\left[-\frac{\sigma}{3\alpha}r^2-\frac{n_0}{r}
+\frac{2f_0^2(5\zeta+14\eta)}{\alpha r^2}\right]^2}{-\frac{2\sigma}{3\alpha}r^2+\frac{4f_0\xi }{\alpha}+\frac{4n_0}{r}
-\frac{4f_0^2(5\zeta+14\eta)}{\alpha r^2}}\right\}.
\label{KS}
\end{align}
Since \eqref{Kre} is independent of the shift function $n(r)/f_{0}$,
the singularity
at $r=0$ originated from the warp factor in front of $dz^{2}$.
When $\sigma=\zeta=\eta=0$, the leading singular behavior at
short distance is $K_{ij}K^{ij}\sim n_{0}/r^{3}$ which matches $1/r^{6}$
behavior in \eqref{4Kr}. The subleading singular behavior is
${\cal O}(1/r^{2})$ with proportionality constant $f_{0}\xi/\alpha$, and
it looks consistent with the matter \eqref{-en} in GR.
Note that there exists another singularity at
$r=-n_{0}\alpha/f_{0}\xi$ from the last term of
\eqref{KS}, only when a negative $n_{0}$ is turned on
in \eqref{KS}~\cite{Cai:2010ud}.

\subsection{Pattern of RG flows}\label{ss22}

\begin{figure*}[ht]\centering
\vspace{10mm}
\scalebox{1.2}[1.2]{
\includegraphics[width=70mm]{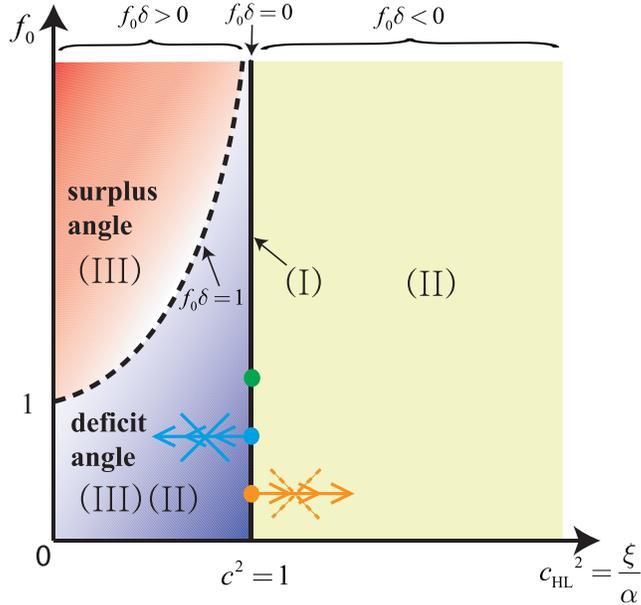}
}\\
\caption{\small Parameter space of $(\xi/\alpha , f_{0})$. The red region (upper left) of
$f_{0}\delta>1$ supports excess cones of surplus angle. As the red color
becomes thicker, the surplus angle increases to $2\pi$.
The dashed curve of $f_{0}\delta=1$ supports a geometry without
deficit/surplus angle. The blue color region (bottom left) of $0<f_{0}\delta<1$ supports
cones with deficit angle. As the blue color becomes thicker, the deficit angle
increases to $2\pi$. The thick vertical line of $\delta=0$ denotes the IR
fixed point of GR. The ivory region (right) of $\delta<0$ involves the geometry of
AdS black string and/or highly curved inner space bounded by a horizon.
}
\label{fig02}
\end{figure*}
From now on let us examine the obtained exact solutions \eqref{mev} near GR
with $\zeta=\eta=0$ and explore possible implication to RG flow in the
vicinity of the IR fixed point, which satisfies the requirements 1-3 in
Sec.~\ref{sec1}. Among the 4 parameters in IR regime
\eqref{8pa} the ratio between two kinetic terms in \eqref{Lir}, $\lambda$,
is assumed to be fixed as unity since the solutions of consideration \eqref{mev} are obtained under $\lambda=1$.
Previously $\sigma$ was identified with a cosmological constant and is left
to be undetermined except the extremely tiny value of the present Universe.
For $\xi$ and $\alpha$, it is convenient to introduce
the variable Newton's constant $G_{{\rm HL}}$ and the variable speed of
graviton $c_{{\rm HL}}$
in HL gravity, visualizing easily the variation from their GR values,
$G$ and $c=1$, as
\begin{align}
\frac{1}{\alpha\xi}=(16\pi G_{{\rm HL}})^{2},\qquad
\frac{\xi}{\alpha}=c_{{\rm HL}}^{2}.
\label{vcg}
\end{align}
The parameter $\alpha$ in front of the kinetic terms is kept to be positive
in order not to make the gravitons the ghosts and then \eqref{vcg} forces
$\xi$ to be non-negative. When $c_{{\rm HL}}^{2}<1$, we have three solutions of
a cone with deficit angle for $f_{0}\delta<1$
(the region of blue color in Fig.~\ref{fig02}),
an excess cone with surplus angle for $f_{0}\delta>1$
(the region of red color), and the space
with zero deficit/surplus angle for $f_{0}\delta=1$ (the boundary of
the dashed line). As discussed previously all three solutions are obtained
without matter in HL gravity \eqref{veq}
but can be obtained with negative energy density
in GR \eqref{-en}.
From the discussion in Sec.~\ref{sec1}, these solutions are
subject to category (III).
Even though a conic geometry with deficit angle can be obtained in GR
with a thin filament type cosmic string of positive energy density,
the large deficit angle around
$2\pi$ has never been observed, which corresponds to $f_{0}\delta\rightarrow
0^{+}$. Since the flow should start from the IR fixed point of GR with
$\delta=0$ (the vertical line of $c_{{\rm HL}}^{2}=c^{2}=1$ in
Fig.~\ref{fig02}),
this cone solution belongs to category (II).
Therefore, we may say that the RG flow to the direction of decreasing the
graviton speed $c_{{\rm HL}}^{2}<1$ (the flow line of sky blue color
in Fig.~\ref{fig02}) seems unlikely and the region of $c_{{\rm HL}}^{2}<1$
may belong to an unlikely zone (the regions of red and blue color in
Fig.~\ref{fig02}).
When $c_{{\rm HL}}^{2}>1$, the constant piece of metric as in \eqref{Em22} can
change signature of the time and radial coordinates. It does not occur
either in the large distance region of the AdS space
$(\sigma>0)$ or in the short distance region of negative mass $(n_{0}<0)$.
Though both the outer space of AdS black string and the highly curved inner
space bounded by a horizon due to a negative mass are allowable and even familiar
theoretically, they have also been hardly detected in the astrophysical observations.
Therefore, the region of $c_{{\rm HL}}^{2}>1$ also seems to be a candidate of
the disfavored zone (II) (the region of ivory color in Fig.~\ref{fig02}) and
the RG flow may not easily run to this direction (the flow line of orange color).
In synthesis,
a possible tentative conclusion in the basis of a few tests favors an RG flow
in which the graviton speed $c_{{\rm HL}}$ remains to be unity
(the dot of green color in Fig.~\ref{fig02}) but the Newton's constant
$G_{{\rm HL}}$ varies like
$16\pi G_{{\rm HL}}=c_{{\rm HL}}/\xi=1/c_{{\rm HL}}\alpha$.
The direction of flow to the regime of weak or strong gravitational coupling
cannot be determined by the classical solutions of consideration \eqref{mev}.

Two comments are in order:
First, once $\xi/\alpha$ flows away from $c^2=1$ before developing other
parameters $\zeta$ and $\eta$ with keeping $\lambda$ to be unity,
the corresponding pure HL gravity can possess a formal general coordinate
invariance based on the Lorentz symmetry with the changed graviton speed
$c_{{\rm HL}}=\sqrt{\xi/\alpha}\neq 1$.
This kind of accidental Lorentz symmetry can also
arise in condensed matter systems, {\it e.g.}, the low energy effective
theory of monolayer graphene with the Fermi
speed $v_{{\rm F}}\sim c/300$~\cite{Semenoff:1984dq}.
Second, near the presumed IR fixed point of GR, both $\xi/\alpha$ and
$\lambda$ parameters in \eqref{Lir} can run away from unity in general.
In this paper, an analysis has always been made by using the solutions with
$\lambda=1$ but assumed $\xi/\alpha$ arbitrary so that its application
to the IR behavior of RG flow is restricted.
Since the structure of highest derivative term in UV regime is based on
the detailed balance with the Cotton tensor, an attractive value of $\lambda$
in the UV fixed point is $\lambda=1/3$ by assuming anisotropic Weyl
symmetry. It may imply a preferred flow path in
which $\lambda$ starts from unity, may decrease monotonic, and
approaches slowly $1/3$.

\setcounter{equation}{0}
\section{Charged Black Holes and Flow for Curvature Square Terms}\label{sec3}

In this section let us take into account both nonvanishing matter
distributions, ${\cal L}_{{\rm M}}\ne 0$, and higher order curvature terms,
$\zeta \neq 0$ and $\eta \neq 0$. We look into their effect from static
axially symmetric solutions and discuss similarity between the effect of
higher derivative terms and that of the static electric field.

Even though we choose $\lambda$ to be unity \eqref{l1}, Eqs.
\eqref{neq3}--\eqref{heq3} are still complicated for nonzero
$\zeta$ and $\eta$, and thus we again restrict
our interest to the solution of constant $f$ \eqref{fcon}.
When we examine various static stringy matter configurations
compatible with the axially symmetric Painlev$\acute{\rm e}$-Gullstrand type metric \eqref{met5},
the corresponding Lagrangian densities for many of those are independent
of the metric function $n(r)$,
$\frac{\partial{\cal L}_{{\rm M}}}{\partial n}=0$,
and then the last matter term on the right-hand side of \eqref{neq3}
vanishes.
It means that
the obtained logarithmic solution of $h$ in \eqref{hsol} with vanishing
integration constant $r_{0}=0$ can also be a solution with the matter.
Insertion of $f$ \eqref{fcon} and $h$ \eqref{hsol} into the remaining
equations \eqref{feq3}--\eqref{heq3} greatly simplifies and reexpresses
the equations \eqref{feq3}--\eqref{heq3} as
a first order equation for $n$,
\begin{align}
\left(rn^2\right)'=\frac{f_0^2\left(5\zeta+14\eta\right)}{\alpha r^2}
+f_{0}\frac{\xi}{\alpha} -\frac{\sigma}{2\alpha} r^2-\frac{r^2}{2\alpha}
\left({\cal L}_{\rm M}-\frac{\partial {\cal L}_{\rm M}}{\partial f}\right),
\label{ne2}
\end{align}
and a first order constraint equation for the matter field,
\begin{align}
\left({\cal L}_{\rm M}-\frac{\partial {\cal L}_{\rm M}}{\partial f}\right)'
=\frac{2}{r}\left(\frac{\partial {\cal L}_{\rm M}}{\partial f}
+\frac{\partial {\cal L}_{\rm M}}{\partial h}\right)
\label{con2}
\end{align}
which should be consistent with the second order Euler-Lagrange equation from
${\cal L}_{{\rm M}}$.

\subsection{Exact solution}\label{ss31}

As a simple but representative example
we consider the electric field produced by
an electrically charged thin filament stretched straightly
along the $z$-direction. Its dynamics is described by the Lagrangian
density
\begin{align}
{\cal L}_{\rm M}=\frac{1}{2}e^{2f}F_{r0}^2.
\label{LE}
\end{align}
For the field theories with anisotropic scaling in their UV regime,
the potential term develops naturally
higher spatial derivative terms but a quadratic time derivative
structure of the kinetic term, $E_{i}^{2}=F_{i0}^{2}$, is unaltered along the flow from IR to UV, which is a good property as a candidate of matter field
in the context of HL gravity.
The profile of the electric field is obtained by solving
Gauss's law with an electric source of charge density per unit length $q_{{\rm e}}$,
\begin{align}
E_r=F_{r0}=e^{-(f+h)}\frac{q_{\rm e}}{4\pi r}.
\label{Erf}
\end{align}
The obtained configurations of $f$ \eqref{fcon}, $h$ \eqref{hsol}, and
$F_{r0}$ \eqref{Erf} satisfy the constraint \eqref{con2}.
In the presence of the electrostatic field we find an exact solution by solving
the remaining equation \eqref{ne2},
\begin{align}
n(r)=\pm \sqrt{-\frac{\sigma}{6\alpha}r^2
+f_0\frac{\xi}{\alpha}+\frac{n_0}{r}-\frac{f_0^2(5\zeta+14\eta)}{\alpha r^2}
-\frac{ q_{\rm e}^2}{64\alpha\pi^2 r^2}}\, .
\label{mee}
\end{align}

If we directly solve the Einstein equations with the same electric field
in the Poincar\'{e} coordinates, we have
\begin{align}
ds^{2}=-\left(-\frac{\Lambda}{3}r^{2}-\frac{GM}{r}
+\frac{G q_{\rm e}^2}{4\pi r^2}\right)
d{\tilde t}^{2}+\frac{dr^{2}}{-\frac{\Lambda}{3}r^{2}-\frac{GM}{r}
+\frac{G q_{\rm e}^2}{4\pi r^2}}+r^{2}(d\theta^{2}+dz^{2}).
\label{GRe}
\end{align}
To compare we perform a coordinate transformation \eqref{tr1} and then obtain
\begin{align}
ds^2=-dt^2+\frac{1}{F_0}\left(dr\pm\sqrt{\frac{\Lambda}{3}r^2
+F_0+\frac{GM}{r}-\frac{G q_{\rm e}^2}{4\pi r^2}}\,\,dt\right)^2
+r^2\left(d\theta^2+dz^2\right).
\label{Em4}
\end{align}
As discussed in the previous section, $\xi/\alpha$ in \eqref{mee} should be
chosen as unity
in the GR limit and, with the help of relation \eqref{par}, the $1/r^{2}$ term
coincides exactly with the term from the electric field in \eqref{GRe} and
\eqref{Em4}.

Since we already discussed physics of the other terms with $\sigma, \xi/\alpha, f_0$, and $n_0
$ in the previous section, we focus  on
the electric field term with $q_{\rm e}^2$ and the higher spatial
derivative terms with $\zeta$ and $\eta$ in \eqref{mee}.
At the level of classical metric \eqref{mee}
both terms are not distinguishable due to
the same $1/r^{2}$ behavior so that it may be natural to introduce an
effective electric charge $q_{{\rm eff}}$ as
\begin{align}
q_{{\rm eff}}^{2}=q_{{\rm e}}^{2}+64\pi^{2}f_{0}^{2}(5\zeta+14\eta).
\label{qeff}
\end{align}
This is manifested in the curvatures: the three-dimensional analog of the Kretschmann invariant is unchanged \eqref{Krev} and the square of the extrinsic curvature keeps its form \eqref{KS} except for a replacement, $f_0^2(5\zeta+14\eta)\rightarrow q_{{\rm eff}}^{2}/64\pi^2$, which means the leading behavior of singularity at the origin is governed by the effective charge \eqref{qeff}.

When $q_{{\rm eff}}^{2}>0$ or equivalently $5\zeta+14\eta>-q_{{\rm e}}^{2}/
64\pi^{2}f_{0}^{2}$, an AdS RN type charged black string can usually be
formed. It may be timely to investigate the horizon in \eqref{mee} in comparison
with \eqref{Em4}.
We assume positive $n_0$ in relation with the positivity of mass in the GR limit.
In AdS space with $\sigma>0$ the possible number of horizons are from zero to two
for $1-\xi/\alpha>0$. In the limit of vanishing cosmological constant
$\sigma\rightarrow 0^{+}$ the horizon is explicitly given as
\begin{align}
\begin{cases}
\displaystyle{ r_{{\rm
\pm}}=\frac{n_0}{2f_0(1-\frac{\xi}{\alpha})}\left[1\pm\sqrt{1-\frac{f_0 q_{{\rm eff}}^{2}}{ 16\alpha\pi^{2}n_0^{2}}\left(1-\frac{\xi}{\alpha}\right)}\,\right]},&
\mbox{when}\quad \displaystyle{\frac{f_0 q_{{\rm
eff}}^{2}}{16\alpha\pi^{2}n_0^{2}}\left(1-\frac{\xi}{\alpha}\right)<1
}\\
\displaystyle{ r_{{\rm e}}=\frac{n_0}{2f_0(1-\frac{\xi}{\alpha})}}, & \mbox{when}\quad
\displaystyle{\frac{f_0 q_{{\rm
eff}}^{2}}{16\alpha\pi^{2}n_0^{2}}\left(1-\frac{\xi}{\alpha}\right)=1}
\\
\mbox{no~horizon}, & \mbox{when}\quad \displaystyle{\frac{f_0 q_{{\rm
eff}}^{2}}{16\alpha\pi^{2}n_0^{2}}\left(1-\frac{\xi}{\alpha}\right)>1 }
\end{cases}
\, .
\label{Hor}
\end{align}
The horizon at $r_{-}$ is analogous to the additional horizon in the
RN black hole, and $r_{\rm e}$ is the unique horizon in the extremal limit.
When $1-\xi/\alpha \rightarrow 0$, $r_{+}$ in \eqref{Hor} moves to infinity and
a single horizon is left at
\begin{align}
r_{{\rm
H}}=\frac{q_{\rm eff}^2}{64\alpha \pi^2 n_0}.
\end{align}
The above charged (black) string structure is maintained even in the absence of
the electric charge $q_{{\rm e}}=0$ as long as $5\zeta+14\eta$ is positive.

When $q_{{\rm eff}}^{2}<0$, this effective charge does not produce an additional horizon but modifies the value of horizon due to mass term $n_0$. When $\sigma \rightarrow 0^{+}$ and $1-\xi/\alpha >0$, it has
\begin{align}
r_{{\rm
H}}=\frac{n_0}{2f_0(1-\frac{\xi}{\alpha})}\left[1+\sqrt{1-\frac{f_0 q_{{\rm eff}}^{2}}{ 16\alpha\pi^{2}n_0^{2}}\left(1-\frac{\xi}{\alpha}\right)}\,\right].
\end{align}
Let us try to obtain the solution in the context of GR under the metric
assumption \eqref{mn}. For vanishing cosmological constant, the Einstein
equations reduce to \eqref{Ett}. Comparing it with \eqref{ne2} in the limit
of $\xi/\alpha=1$, $\sigma=0$, and ${\cal L}_{{\rm M}}=0$, we have
$8\pi G (-T^{t}_{\; t})=f_{0}^{2}(5\zeta+14\eta)/\alpha r^{4}$ which leads to
a distribution of negative energy density for $5\zeta+14\eta<0$.
Obtaining this solution in GR violates the positive energy theorem, but
it is a generic vacuum solution in HL gravity.
In case that the source is assumed to be an electrostatic field, $E_{r}$,
then it requires $-T^{t}_{\; t}= f_{0}E_{r}^{2}/2<0$. Substituting the solutions
\eqref{fcon} and \eqref{hsol} into \eqref{Erf}, the energy density has
$-T^{t}_{\; t}=q_{{\rm e}}^{2}/32\pi^{2} r^{4}$. Therefore, we read
the effect of quartic spatial derivative terms as that of
electric charge $q_{{\rm e}}^{2}=4\pi f_{0}^{2}(5\zeta+14\eta)/G\alpha$ and
then $q_{{\rm e}}^{2}$ can be interpreted to be negative for $5\zeta+14\eta<0$.
It is forbidden to let the square of physical electric charge negative,
$q_{{\rm e}}^{2}< 0$ in Maxwell theory,
since it means that the classical Coulomb force
is attractive between two electric charges of the same sign.
The square of effective charge \eqref{qeff},
on the other hand, can take a negative value for
$5\zeta+14\eta < -q_{{\rm e}}^{2}/64\pi^{2}f_{0}^{2}$.
Since the kinetic term given by the square of the electric
field~\cite{Horava:2008jf} is kept to be quadratic,
the property of attractive Coulomb force between the same charges holds
in the Lifshitz type field theories irrespective of the
anisotropic scaling in the matter sector.

Since the $1/r^{2}$ terms from both the higher spatial
derivative terms and electric charges are subleading to the $1/r$ term
in the astronomical scale, it is difficult to
detect such a feeble astrophysical signal in the long distance physics.
Here we have some qualitatively unusual results in the context of GR so it
may be worth studying their astrophysical effect originated from
indistinguishability between the higher derivative term effect and the electric
charge.
Astrophysical realization of the charged black holes and strings
is believed to be
difficult since electric charges repel each other. In the UV regime
of HL gravity with higher spatial derivative terms of positive
$5\zeta+14\eta$, production of the
charged black strings seems not to be hindered by the repulsive nature
between electric charges. Suppose such charged black strings have
sufficiently been
generated in the early universe due to the quartic spatial derivative terms and
keep their nature along the flow to
the present IR regime. Although the assumptions seem very rough, such
objects or their indirect astrophysical signals may be observed
occasionally in the present universe. In this sense future black hole
observations including their detailed properties may provide some possibility
to be used to constrain
 the possible parameter region of $5\zeta+14\eta$ as long as one cannot find
a compelling reason to set $f_0$ to be sufficiently small.

In this section we assumed the same electric field of accumulated charges, and
the obtained results depend on the form of metric and the assigned conditions
in HL gravity. In the previous work~\cite{Kim:2009dq}, on the other hand,
the detailed balance condition was
assumed in the entire energy scale but the projectability condition was not.
As a result, the electric field from accumulated charges induced a geometry with
deficit or surplus angle for $1/3\le \lambda <1/2$.
In the present work the projectability condition
is assumed but the detailed balance condition is imposed only at the UV
fixed point.
Then the same electric field source leads to the $1/r^{2}$ term in the
metric for $\lambda=1$, which is the same as the usual $1/r^{2}$ term in the RN
type charged black hole.
Between GR and HL gravity, this kind of difference basically originates
from higher spatial derivative terms. It also shows that
HL gravity theories become different with and without the detailed balance
condition, even though they share similar higher spatial derivative terms.

\subsection{Pattern of RG flows}\label{ss32}

\begin{figure*}[ht]\centering
\vspace{10mm}
\scalebox{1.2}[1.2]{
\includegraphics[width=75mm]{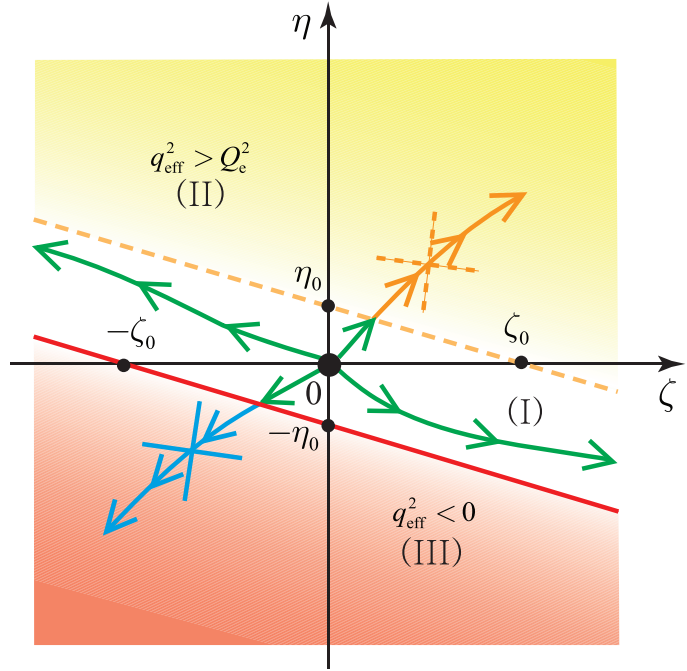}
}\\
\caption{\small Parameter space of $(\zeta , \eta)$.
The nature of Coulomb force between two charges of the same charge is
effectively attractive in the red region (lower) of
$5\zeta+14\eta<-Q_{{\rm e}}^{2}/64\pi^{2}f_{0}^{2}$.
The size of the effective charge is larger than the upper bound
$|Q_{{\rm e}}|$ in the yellow region (upper) of
$5\zeta+14\eta>Q_{{\rm e}}^{2}/64\pi^{2}f_{0}^{2}$.
Here $\zeta_{0}=Q_{{\rm e}}^{2}/896\pi^{2}f_{0}^{2}$ and
$\eta_{0}=Q_{{\rm e}}^{2}/320\pi^{2}f_{0}^{2}$.
}
\label{fig03}
\end{figure*}
Discussion on the patterns of RG flows utilizing the classical solutions
becomes subtle for the coefficients of quartic spatial derivative terms,
$\zeta$ and $\eta$ of $R_{ij}R^{ij}$ and $R^{2}$ terms respectively.
If nonzero values of $\zeta$ and $\eta$ are developed just above the Planck
scale near the IR fixed point, application of the classical solutions may be
more reliable. If those coefficients start to develop their nonzero values
at some intermediate energy scale which is much larger than the Planck scale but
lower than the scale of the Lifshitz fixed point, it is skeptical about the
scenario relying heavily on the GR data, theoretical and observational.
Coulomb repulsion among the charges of the same sign disfavors the formation
of charged astronomical objects.
Therefore, we frequently observe neutral objects like almost all the stars and
neutron stars.
Though the charge of RN type black holes
can arbitrarily be large, accumulation of the excess
charge by the physical process is probably limited by an extremely tiny
ratio to mass, and thus, in a statistical sense, we can naturally
assume an upper bound of the accumulated charge $Q_{{\rm e}}$.
As discussed in Sec.~\ref{ss31}, the square of the effective charge can be
negative, $q_{{\rm eff}}^{2}<0$, despite positive $q_{{\rm e}}^{2}$.
Therefore the effect of quartic spatial derivative terms corresponds to the
attractive Coulomb force between two charges of the same sign
and introduction of negative energy density in GR, which are not allowed
in electrodynamics and in GR, respectively.
According to the forbidden attractive Coulomb force between two charges
of the same sign and violation of the positive energy theorem in GR,
the solution belongs to category (III) and
the region of negative $q_{{\rm eff}}^{2}$ may be an unlikely zone
for the RG flows for $\zeta$ and $\eta$,
$5\zeta+14\eta<-Q_{{\rm e}}^{2}/64\pi^{2}f_{0}^{2}$ (the region of red color
and the flow line of sky blue color in Fig.~\ref{fig03}).
Since an object carrying the electric charge larger
than $Q_{{\rm e}}$ can rarely be observed, the solution belongs to
category (II) and the entrance of RG flows
to the region of $5\zeta+14\eta>Q_{{\rm e}}^{2}/64\pi^{2}f_{0}^{2}$ may also
be disfavored (the region of yellow color and the flow line of orange color
in Fig.~\ref{fig03}).
Therefore, the RG flows about the line of $5\zeta+14\eta=0$ are allowed
(the band of white color and the flow lines of green color in
Fig.~\ref{fig03}). The value of upper bound $Q_{{\rm e}}$ is not
estimated in the absence of matter-sector physics of the field
theories with the Lifshitz point, so is not the width of the white band.

\subsection{Exact solutions with matters of power-law behavior}\label{ss33}

Next let us assume some matter distribution showing the following power-law behavior
\begin{align}
{\cal L}_{\rm M}-\frac{\partial {\cal L}_{\rm M}}{\partial f}=\frac{A}{r^a},\qquad {\cal L}_{\rm M}+\frac{\partial {\cal L}_{\rm M}}{\partial h}=\frac{B}{r^b},
\label{Lm}
\end{align}
where $A$ and $B$ are undetermined constants. Substituting the configuration \eqref{Lm} into the
constraint equation \eqref{con2}, we obtain $a=b$ and $B=(1-a/2)A$.
With these relations  we solve Eq. \eqref{ne2} exactly and find the
solutions
\begin{align}
n(r)=
\begin{cases}
\displaystyle{\pm \sqrt{-\frac{\sigma}{6\alpha}r^2+\frac{f_0\xi}{\alpha}+\frac{n_0}{r}-\frac{f_0^2(5\zeta+14\eta)}{\alpha r^2}-\frac{A\, r^{2-a}}{2\alpha(3-a)}}}\,\,,&\quad (a \geq 0,~a \neq 3) \\
\displaystyle{\pm\sqrt{-\frac{\sigma}{6\alpha}r^2+\frac{f_0\xi}{\alpha}
+\frac{n_0}{r}-\frac{f_0^2(5\zeta+14\eta)}{\alpha r^2}-\frac{A \ln r}{2\alpha r}}}\,\,,& \quad(a=3)
\end{cases}
.
\end{align}

Note that the $a=4$ case is nothing but the aforementioned electric field of charge density $q_{\rm e}$ with $ A=-q_{\rm e}^2/32\pi^2$ and $B=q_{\rm e}^2/32\pi^2$.
A constant matter distribution, the $a=0$ case, affects the cosmological constant. According to power counting, the physics of odd $a$ is unclear as was discussed in the previous work~\cite{Kim:2009dq}. For the magnetic field of a thin filament
of magnetic charge density $q_{\rm m}$ in IR regime ($z=1$),
we can obtain the same solution \eqref{mee} except for the replacement
$q_{\rm e}\rightarrow q_{\rm m}$ since the electromagnetic duality of
Maxwell theory works. The cases of even $a$'s maybe interpreted as the
magnetically charged thin filament with different anisotropic scalings of
$z=a/2-1$. To be a consistent solution, the field configuration of
consideration should simultaneously satisfy both the constraint equation
\eqref{con2} and the Euler-Lagrange equation from ${\cal L}_{\rm M}$.

An intriguing example of a stringy object showing a power-law behavior in its
energy density may be global vortices. The gravitating global vortex in the IR regime
of $z=1$ is described by a field theory of a complex scalar field $\phi$ and
the Lagrangian density for the $k$ vortices superimposed at the origin is
\begin{align}
{\cal L}_{\rm M}=-\frac{1}{2}e^{2f}|\phi|'^2
-\frac{k^2}{2r^2}|\phi|^2
-V(|\phi|), \label{Lsc}
\end{align}
where $V(|\phi|)$ is an arbitrary scalar potential of scalar amplitude
$|\phi|$ involving spontaneous symmetry breaking,
$|\phi|\stackrel{r \rightarrow \infty}{\rightarrow}v$.
It is then natural to ask whether or not the $1/r^2$ term for nonzero
vorticity $(k\neq0)$ leads to
a possible $a=2$ solution. In order to be compatible with the metric solution
in \eqref{fcon} and \eqref{hsol}, the scalar amplitude should satisfy
the constraint equation \eqref{con2} as well as the Euler-Lagrange equation
derived from \eqref{Lsc}. A straightforward analysis proves nonexistence of
the static global vortex solutions described by \eqref{Lsc}.

\setcounter{equation}{0}
\section{Conclusion and Discussion}\label{sec4}

In this paper we have considered $(1+3)$-dimensional HL gravity of $z=3$
anisotropic scaling. We employ the version with the projectability condition
but impose the detailed balance condition only in the UV fixed point, which
results in the Lagrangian containing 8 parameters and the square of the Cotton
term as the unique sixth order spatial derivative term.
Solving the equations of motion in the Painlev\'{e}-Gullstrand type
coordinates with axial symmetry, we found exact static stringy solutions
which allow various planar geometries: a cone with deficit angle, an excess cone
with surplus angle, AdS type black holes, and a charged black hole.
The excess cone with surplus angle is a forbidden solution due to
violation of the positive energy theorem in GR, but, amazingly enough,
it is a generic vacuum solution in HL gravity.
The effect of quartic spatial derivative terms
is equivalent to the square of the electric field in static solution, of which
negative coefficients permit the square of electric field to be negative.
In order to obtain such a
solution in GR the positive energy theorem is violated.
It can also be interpreted as the effect making the square of electric charge
negative, which is against repulsive Coulomb
force between the same charges in Maxwell theory.
The obtained solutions are categorized and tested according to the
classification
stated in Sec.~\ref{sec1}, and the results are summarized in
Figs.~\ref{fig02} and~\ref{fig03} and in Table I.

\vspace{3mm}
\noindent\noindent
\begin{center}{
\begin{tabular}{|c|c|c|c|c|}
\hline\hline Solution & Parameters & Category & Zone & Physics \\
\hline\hline Excess cone & $f_0(1-\xi/\alpha)>1$ & (III) & & \\
\cdashline{1-3}
Vanishing& & & & Violation of \\
deficit/surplus& $f_0(1-\xi/\alpha)=1$ & (III) & Unlikely & the positive energy \\
angle & & & & theorem in GR \\
\cdashline{1-3}
Cone & $0<f_0(1-\xi/\alpha)<1$ & (III), (II) & & \\
\hline   GR limit & $f_0(1-\xi/\alpha)=0$ & (I) & Allowed &
Equivalent to GR solution\\
\hline & & & & Spacetime signature flip\\
AdS  & $f_0(1-\xi/\alpha)<0$ & (II) & Disfavored & without a negative \\
black string& & & & cosmological constant\\
\hline\hline
& & & &Attractive Coulomb\\
& $5\zeta+14\eta<-Q_{\rm e}^2/64\pi^2f_0^2$ & (III) & Unlikely & force between \\
Charged& & & & the same charges \\
\cline{2-5}(black) & $5\zeta+14\eta\approx 0$ & (I) & Allowed &
Effectively (almost) neutral\\
\cline{2-5}
string& & & &Square of effective \\
& $5\zeta+14\eta>Q_{\rm e}^2/64\pi^2f_0^2$ & (II) & Disfavored &
excess charge is positive \\
& & & & but too large\\
\hline\hline
\end{tabular}
}

\vspace{0mm}
{\small Table 1. Classification of solutions and the categorized zones}
\end{center}

The second purpose of this paper was to
propose a method to constrain a possible pattern of the RG flows connecting
the IR and UV fixed points by utilizing generic classical solutions of HL
gravity. Since we cannot directly tackle the quantum regime of HL gravity
and cannot compute the RG flows at the moment, this indirect method we propose
may provide
some hints on viable zones in the space of many parameters.
Specifically, in Sec.~\ref{sec1}, we suggested the guidelines for
selecting classical solutions, 3 categories, and the classification scheme
of the parameter regions, 3 zones.
However, probing the quantum nature of HL gravity by using classical tools
involves ambiguities and jumps.
Though we have some generic classical solutions
in exact form, some theoretical arbitrariness always exists in categorizing
the solutions and selecting the zones. In addition,
survival of the corresponding objects and detection of their
astrophysical signal in the present universe are endangered by the following
possibilities. Above all, their production rate can be low.
Even if the objects are materialized in the very early universe,
they can be swept away by environmental changes, can be recombined to some
different objects by the mutual interactions,
can be transformed to fossils in the period of phase
transition, and can be diluted in the inflationary era.
All of these lead to a rare event rate or feeble signals
in astronomical observations.
However, if the obtained solutions are generic and the corresponding objects are
formed in the epoch of HL gravity,
they are likely inherited and their remnants might be yet residing in
the present universe.

The above ambiguities of the conclusion on the parameter zones, obtained from
a few tests based on the exact classical solutions, are tentative at this moment.
Subsequently this indistinct conclusion in the beginning stage even obscures
validity of the proposed method itself. However, fortunately enough,
many classical solutions, among which some are generic and exact, have already
been found in HL gravity, and more classical
configurations, of which characteristics are drastically different from
GR solutions, will be obtained. If these many solutions are applied for
attaining a viable pattern of the RG flows and for testing the proposed method,
we will have various maps filled with classified zones and their boundaries as
given in Figs.~\ref{fig02} and \ref{fig03}. Then we overlap these maps
on each corresponding lower-dimensional scene projected from
the multidimensional parameter
space. For example, Fig.~\ref{fig02} (Fig.~\ref{fig03}) is understood as
a (another) map to put on a (a different) 2-dimensional scene projected
from our 10-dimensional space of 8 parameters and 2 constants.
Once the overlapped data are sufficiently accumulated, the information
on every zone becomes manifest, classified by the allowed, disfavored,
and unlikely zone that we explained earlier. Then a panoramic view of
the whole multidimensional parameter space is also available.
This result will suggest a possible pattern of the RG flows, and, if we are
lucky enough, will provide some restricted narrow paths for the flows.
When we reach a consistent conclusion on the pattern of the RG flows
through the tests utilizing classical solutions, it will help to judge
whether or not there can exist a viable flow line
connecting smoothly the UV fixed point of $z=3$ anisotropic scaling and
the IR fixed point of GR.
Furthermore, once the aforementioned scenario is proven to be
reliable, the panoramic method we propose will be established
as an efficient way in studying the RG flows of a premature complicated
theory involving too many terms and undetermined coefficients,
and finally will play a role of {\it classical} predictor for unknown
{\it quantum} realm.

We conclude this section with a list of further studies.
First, we obtained and tested the static solutions with axial symmetry, but
it is also worth finding those with spherical symmetry for additional
tests~\cite{Kim:2010vs}.
Second, the exact solutions are obtained only for $\lambda=1$.
If we can find some exact solutions for arbitrary $\lambda$ despite
extreme complication,
it will allow a more accurate discussion on the RG flow in the IR limit.
Third, our studies focused on the IR regime.
To understand quantized HL gravity it is important to figure out
physics in the vicinity of the UV fixed point, where higher spatial derivatives
satisfying the $z=3$ anisotropic scaling dominate.
For the application of the proposed panoramic method,
it is intriguing
to find generic classical solutions of the nonvanishing Cotton term $C_{ij}\ne 0$.
This study in UV regime may also shed light on understanding the phase
structure of HL gravity, which is performed only through the comparison
to lattice results~\cite{Xu:2010eg}.

\section*{Acknowledgments}

The authors would like to thank Cheongho Han, Hang Bae Kim, and Sung-Soo Kim
for valuable discussions and comments.
This work was supported by the National Research Foundation of Korea(NRF)
grant funded by the Korea government (MEST) (No. 2009-0062869) through the
Astrophysical Research Center for the Structure and Evolution of the Cosmos
(ARCSEC) and
by the Korea Research Foundation Grant funded by the Korean Government
(KRF-2008-313-C00170).

\end{document}